\documentclass[11pt]{article}
\usepackage[pctex32]{graphics}

\def\ben{\begin{equation}}
\def\een{\end{equation}}

\def\nn{\nonumber} \def\bd{\begin{document}} \def\ed{\end{document}}
\def\ds{\documentstyle} \let\fr=\frac \let\bl=\bigl \let\br=\bigr
\let\Br=\Bigr \let\Bl=\Bigl
\let\bm=\bibitem
\let\na=\nabla
\let\pa=\partial \let\ov=\overline
\newcommand{\be}{\begin{equation}}
\newcommand{\ee}{\end{equation}}
\def\ba{\begin{array}}
\def\ea{\end{array}}
\def\ft#1#2{{\textstyle{\frac{\scriptstyle #1}{\scriptstyle #2} } }}
\def\fft#1#2{{\frac{#1}{#2}}}
\def\del{\partial}
\def\vp{\varphi}
\def\sst#1{{\scriptscriptstyle #1}}
\def\oneone{\rlap 1\mkern4mu{\rm l}}
\def\td{\tilde}
\def\wtd{\widetilde}
\def\ie{{\it i.e.\ }}
\def\dalemb#1#2{{\vbox{\hrule height .#2pt
        \hbox{\vrule width.#2pt height#1pt \kern#1pt
                \vrule width.#2pt}
        \hrule height.#2pt}}}
\def\square{\mathord{\dalemb{6.8}{7}\hbox{\hskip1pt}}}
\newcommand{\ho}[1]{$\, ^{#1}$}
\newcommand{\hoch}[1]{$\, ^{#1}$}
\newcommand{\bea}{\setlength\arraycolsep{2pt} \begin{eqnarray}}
\newcommand{\eea}{\end{eqnarray}}
\newcommand{\ra}{\rightarrow}
\newcommand{\lra}{\longrightarrow}
\newcommand{\Lra}{\Leftrightarrow}
\newcommand{\bp}{\tilde \beta^\prime}
\newcommand{\tr}{{\rm tr} }
\newcommand{\Tr}{{\rm Tr} }
\def\0{{\sst{(0)}}}
\def\1{{\sst{(1)}}}
\def\2{{\sst{(2)}}}
\def\3{{\sst{(3)}}}
\def\4{{\sst{(4)}}}
\def\5{{\sst{(5)}}}
\def\6{{\sst{(6)}}}
\def\7{{\sst{(7)}}}
\def\8{{\sst{(8)}}}
\def\m{{\sst{(m)}}}
\def\n{{\sst{(n)}}}
\def\cA{{{\cal A}}}
\def\cB{{{\cal B}}}
\def\cF{{{\cal F}}}
\def\cG{{{\cal G}}}
\def\cH{{{\cal H}}}
\def\tV{\widetilde V}
\def\tW{\widetilde W}
\def\tH{\widetilde H}
\def\tE{\widetilde E}
\def\tF{\widetilde F}
\def\tA{\widetilde A}
\def\im{{{\rm i}}}
\def\tY{{{\wtd Y}}}
\def\ep{{\epsilon}}
\def\vep{{\varepsilon}}
\def\bD{{{\bar D}}}
\def\R{{{\mathbb R}}}
\def\C{{{\mathbb C}}}
\def\H{{{\mathbb H}}}
\def\CP{{{\mathbb C}{\mathbb P}}}
\def\RP{{{\mathbb R}{\mathbb P}}}
\def\Z{{{\mathbb Z}}}
\def\bA{{{\mathbb A}}}
\def\bB{{{\mathbb B}}}
\def\bC{{{\mathbb C}}}
\def\bD{{{\mathbb D}}}
\def\bE{{{\mathbb E}}}
\def\bZ{{{\mathbb Z}}}
\def\Re{{{\frak{Re}}}}
\def\Im{{{\frak{Im}}}}
\def\cosec{{\,\hbox{cosec}\,}}
\def\Gm{{\Gamma_{\!\! -}}}
\def\Gp{{\Gamma_{\!\! +}}}
\def\stan{{standard }}
\def\nonstan{{supernumerary }}
\def\p{{\partial}}
\def\kdel#1{{\fft{\del}{\del#1}}}

\def\bog{{Bogomolny }}
\def\om{{\omega}}

\newcommand{\nnr}{\nonumber \\}
\newcommand{\pd}{\partial}
\newcommand{\ud}{\textrm{d}}
\newcommand{\dTH}{T^{\prime \, 0}_\textrm{H}}
\newcommand{\dOi}{\Omega^{\prime \, 0}_i}
\newcommand{\bx}{{\bf x}}

\begin{document}

\vspace{5mm}
\begin{center}
{\Large \bf Thermodynamics of black holes in the deformed
Ho\v{r}ava-Lifshitz gravity } \vspace{12mm}

{\large   Yun Soo Myung \footnote{e-mail
 address: ysmyung@inje.ac.kr}}
 \\
\vspace{10mm} {\em Institute of Basic Science and School of
Computer Aided Science \\ Inje University, Gimhae 621-749, Korea}
\end{center}

\begin{center}

\underline{Abstract}
\end{center}

  We study thermodynamics of black holes in the deformed Ho\v{r}ava-Lifshitz gravity  with coupling constant $\lambda$.
  For $\lambda=1$, the black hole behaves the
  Reissner-Norstr\"om black hole. Hence, this is different from the Schwarzschild black
  hole of Einstein gravity. A connection to the generalized
  uncertainty principle is explored to understand the Ho\v{r}ava-Lifshitz black
  holes.

\vspace{15pt}

\thispagestyle{empty}





\newpage
\section{Introduction}
Recently Ho\v{r}ava has proposed a renormalizable theory of
gravity at a Lifshitz point\cite{ho1},  which  may be regarded as
a UV complete candidate for general relativity.  Very recently,
the Ho\v{r}ava-Lifshitz gravity theory has been intensively
investigated in~\cite{ho2,ho3,VW,klu,Nik,Nas,Iza,Vol,CH,CHZ} and
its cosmological applications have been  studied in
~\cite{cal,TS,muk,Bra,pia,gao}.

Introducing the ADM formalism where the metric is parameterized
\cite{adm}
\be ds_{ADM}^2= - N^2  dt^2 + g_{ij} \Big(dx^i - N^i dt\Big)
\Big(dx^j - N^j dt\Big)\,, \ee
the Einstein-Hilbert action can be expressed as
\be \label{Eins} S_{EH} = \fft{1}{16\pi G} \int d^4x \sqrt{g} N
\Big(K_{ij} K^{ij} - K^2 + R - 2\Lambda\Big)\,, \ee
where $G$ is Newton's constant and extrinsic curvature $K_{ij}$
takes the form
\be K_{ij} = \fft{1}{2N} \Big(\dot g_{ij} - \nabla_i N_j -
\nabla_j N_i\Big)\,. \ee
Here, a dot denotes a derivative with respect to $t$.

We would like to mention that the IR vacuum of this theory is anti
de Sitter (AdS) spacetimes. Hence, it is interesting to take a
limit of the theory, which may lead to  a Minkowski vacuum in the
IR sector. To this end, one may modify the theory by introducing
``$\mu^4R$" and then, take the $\Lambda_W \to 0$ limit. This does
not alter the UV properties of the theory, but it changes the IR
properties. That is, there exists a Minkowski vacuum, instead of
an AdS vacuum. Then, one finds the black hole solution, deviating
from the Schwarzschild solution.

 A deformed action of the non-relativistic renormalizable gravitational
theory  is given by~\cite{KS}
\bea%
S_{HL}&=&\int dtd^3\bx\, \Big({\cal L}_0 + \tilde{{\cal L}}_1\Big)\,,\nn\\
{\cal L}_0 &=& \sqrt{g}N\left\{\frac{2}{\kappa^2}(K_{ij}K^{ij}
\label{action1}-\lambda K^2)+\frac{\kappa^2\mu^2(\Lambda_W R
  -3\Lambda_W^2)}{8(1-3\lambda)}\right\}\,,\\ \tilde{{\cal L}}_1&=&
\sqrt{g}N\left\{\frac{\kappa^2\mu^2
(1-4\lambda)}{32(1-3\lambda)}R^2 -\frac{\kappa^2}{2w^4}
\left(C_{ij} -\frac{\mu w^2}{2}R_{ij}\right) \left(C^{ij}
-\frac{\mu w^2}{2}R^{ij}\right) +\mu^4R \right\}\,\label{action2}
\eea%
where $C_{ij}$ is the Cotton tensor
\be
C^{ij}=\epsilon^{ik\ell}\nabla_k\left(R^j{}_\ell
-\frac14R\delta_\ell^j\right)=\epsilon^{ik\ell}\nabla_k R^j{}_\ell
-\frac14\epsilon^{ikj}\pd_kR\,.\label{def.K.C}
\ee
Comparing ${\cal L}_0$ with Eq.(\ref{Eins}) of general relativity,
the speed of light, Newton's constant and the cosmological
constant are given by
\be c=\fft{\kappa^2\mu}{4}
\sqrt{\fft{\Lambda_W}{1-3\lambda}}\,,\qquad
G=\fft{\kappa^2}{32\pi\,c}\,,\qquad \Lambda=\ft32
\Lambda_W\,.\label{cg} \ee
The equations of motion were derived in \cite{LMP} and \cite{KK},
but we do not write  them  due to the length.

\section{Thermodynamics of Ho\v{r}ava-Lifshitz black holes}

In this section, we investigate the  black hole solution in the
limit of $\Lambda_W \to 0$  and its thermodynamic properties. For
this purpose, considering $N^i=0$, the spherically symmetric
solutions could  be obtained with the metric
ansatz~\cite{LMP,CCO,CLS,CY,MK,Nis,CCO1,Gho}
\be \label{ssm} ds_{SS}^2 = - N^2(r)\,dt^2 + \fft{dr^2}{f(r)} +
r^2 (d\theta^2 +\sin^2\theta d\phi^2)\,. \ee
 Considering  the minimal theory, we obtain the
Schwarzschild black hole whose metric function is given by
\be \label{smet} f_S=1 - \fft{2m}{r}\,.\label{adsbh} \ee
 In order to obtain the solution, let us
substitute the metric ansatz (\ref{ssm}) into the action, and then
vary the functions $N$ and $f$. This is possible because the
metric ansatz shows all the allowed singlets which are compatible
with the $SO(3)$ action on the $S^2$. The reduced Lagrangian is
given by
\be \label{react} \tilde{{\cal L}}_1= \fft{ \kappa^2\mu^2 N
}{8(1-3\lambda)\sqrt{f}}\Bigg(  \fft{\lambda-1}{2} f'^2 -
\fft{2\lambda (f-1)}{ r}f' + \fft{(2\lambda-1)(f-1)^2}{
r^2}-2\omega(1 - f - r f')\Bigg) \ee
with $\omega=8\mu^2(3\lambda-1)/\kappa^2$. For
$\lambda=1~(\omega=16\mu^2/\kappa^2)$, we have a solution where
$f$ and $N$ are determined to be \be \label{sol1} N^2=f=1 + \omega
r^2-\sqrt{r(\omega^2 r^3+4\omega m)}\, \label{nf} \ee where $m$ is
an integration constant related to the mass.

\begin{figure}[t!]
   \centering
   \includegraphics{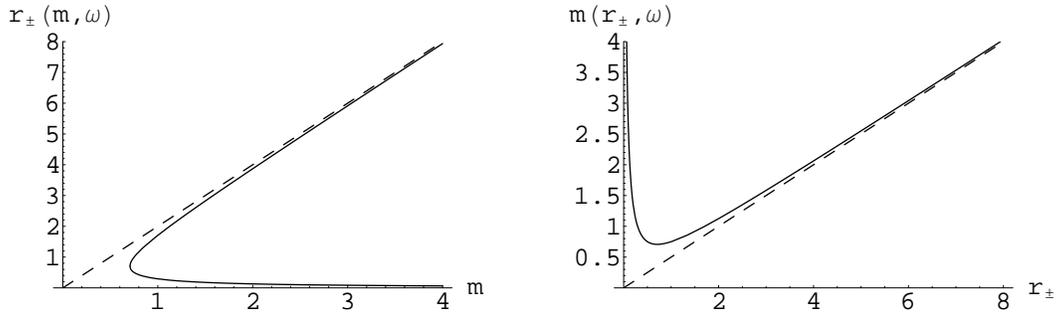}
\caption{Graphs of two horizons ($r_\pm$) and mass $m$ with
$\omega=1$. Left graph: the upper solid curve represents the outer
horizon $r_+(m,\omega)$ and the lower shows the inner horizon
$r_-(m,\omega)$. The dashed line denotes the Schwarzschild black
hole.
 Right graph:
$m(r_\pm,\omega)$ with $\omega=1$. For $r_-<r_e=0.71$, the solid
curve represents $m(r_-,\omega)$ while for $r_+>r_e$, it
represents $m(r_+,\omega)$. The dashed line denotes the
Schwarzschild black hole.} \label{fig.1}
\end{figure}
\begin{figure}[t!]
   \centering
   \includegraphics{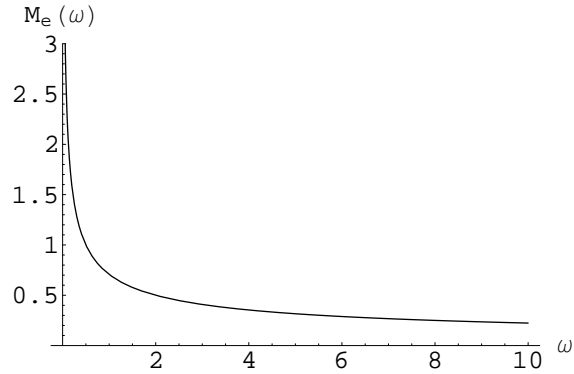}
\caption{Extremal mass graph. The curve denotes the extremal black
hole $m_e(\omega)$. The naked singularity regions are  area below
the curve while the non-extremal black hole are area above the
curve.} \label{fig.2}
\end{figure}

As is shown in Fig. 1, from the condition of $f(r_\pm)=0$, the
outer (inner) horizons are given by \be \label{sol}
r_\pm=m\Bigg[1\pm \sqrt{1-\frac{1}{2\omega m^2}}\Bigg]. \ee In
order to have a black hole solution, it requires that \be m^2 \ge
\frac{1}{2\omega}. \ee Further, the degenerate horizon \be
r_e=m_e=\frac{1}{\sqrt{2\omega}}\ee appears when \be m^2=m_e^2.
\ee A graph of $m_e(\omega)$ is shown in Fig. 2, splitting the
whole region into  the non-extremal black holes and the naked
singularity regions. We do not have such a curve in the
Schwarzschild black hole because the Schwarzschild black hole has
a single horizon only.

 The mass function is defined as \be
m(r_\pm,\omega)=\frac{1+2\omega r^2_\pm}{4\omega r_\pm}. \ee The
mass is depicted in Fig. 1. For large $r_+$, it reduces to the
mass of Schwarzschild black hole ($m_S=r_+/2$).

  The
temperature  is defined  by
\be \label{temp}T=\frac{f'}{4\pi}\mid_{r=r_+}= \frac{2\omega
r_+^2-1}{8\pi(\omega r_+^3+r_+)}, \ee
while for large $r_+$, it leads  to the Schwarzschild case\be
T_{S}=\frac{1}{ 4\pi r_+}. \ee  We confirm that $T=0$ for the
extremal black hole at $r_e=1/\sqrt{2\omega}$. As is depicted in
Fig. 3, we observe that
 the maximum point $r_+=r_m$  appears as
 \be
 x_m=\frac{\sqrt{5\sqrt{33}}}{2\sqrt{\omega}}\ee
which satisfies $dT/dr_+=0$. Hence, for $\lambda=1$, the solution
interpolates between AdS$_2\times S^2$ in the near-horizon
geometry of extremal black hole and M$_4$ (asymptotically flat
spacetimes).
\begin{figure}[t!]
   \centering
   \includegraphics{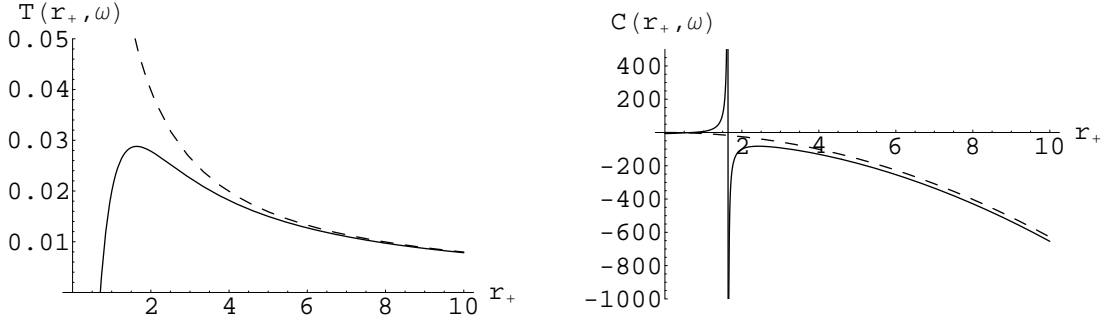}
\caption{Temperature and heat capacity graphs. Left graphs:
$T(r_+,\omega)$  with $\omega=1$. We observe that
$T(r_e,\omega)=0$ at $r_e=0.71$ while $T=T_m$ at $r_m=1.64$. The
dashed curve denotes the temperature $T_S$ of the Schwarzschild
black hole.
 Right
graph: $C(r_+,\omega)$ shows a blow-up point
 at $x_m=1.64$, dividing it into $C>0$ and $C<0$. Note that $C(r_e,\omega)=0$.
 The dashed curve denotes
the heat capacity $C_S$ of the Schwarzschild black hole.}
\label{fig.3}
\end{figure}

Finally, the heat capacity defined by $C=
\Big(\frac{dm}{dT}\Big)_\lambda$ takes the form
 \be C(r_+,\omega)=-\frac{2\pi}{\omega}\frac{(\omega r_+^2+1)^2(2\omega r_+^2-1)}{2\omega^2 r_+^4-5\omega r_+^2-1}. \ee
An isolated black hole like Schwarzschild black hole is never in
thermal equilibrium because it decays by the Hawking radiation. It
could be represented by its negative heat capacity. However, the
Schwarzschild black hole can be in thermal equilibrium in anti-de
Sitter spacetimes (AdS$_4$) because the AdS$_4$ plays the role of
a box. In this case, we have small black hole with negative heat
capacity and large black hole with positive heat capacity.  Hence
one should consider not just the black hole itself, but also its
surroundings of the heat reservoir. Actually, the thermodynamic
stability argument based on the sign of heat capacity is closely
related to the stability argument based on the sign of
eigenfrequency $w^2$ in the perturbation theory~\cite{GM}. Here we
have the correspondence between two stabilities:

\begin{tabular}{|c|c|c|}
  \hline
   & Thermodynamic stability & Classical stability\\
  \hline
  stability & $C>0$ & $w$=real ($w^2>0$) \\
  instability & $C<0$ & $w$=imaginary ($w^2<0$) \\
  \hline
\end{tabular}

 As is shown in Fig. 3,  the heat capacity blows up at $r_+=r_m$.
 Hence, small black holes of $r_+<r_m$ are
stable because  their heat capacities are positive, while large
black hole of $r_+>r_m$ are unstable because their heat capacities
are negative, like the heat capacity $C_S=-2\pi r_+^2$  of
Schwarzschild black hole. In this case, we have positive heat
capacities for  small black holes even in asymptotically flat
spacetime because of the presence of $\omega$, like the charge $Q$
in the Reissner-Nordstr\"om black hole.

\section{GUP and Ho\v{r}ava-Lifshitz gravity}

The solution in Eq.(\ref{sol}) reminds us the Schwarzschild black
hole modified  with the generalized uncertainty principle
(GUP)~\cite{MKP}. Hence, we make
 a connection between the GUP and the
Ho\v{r}ava-Lifshitz gravity as a UV complete candidate for general
relativity.  A commutation relation of
         \begin{equation} \label{crgup}
         [\vec{x}, \vec{p}]=i\hbar(1+\beta^2\vec{p}^2)
         \end{equation}
         leads to the  generalized uncertainty relation
 \be \label{1eq2} \Delta x \Delta p \ge \hbar
\Bigg[1+\alpha^2 l_p^2 \frac{(\Delta p)^2}{\hbar^2}\Bigg] \ee with
$l_p=\sqrt{G \hbar/c^3}$ the Planck length. Here  a parameter
$\alpha=\hbar\sqrt{\beta}/l_p$ is introduced to indicate the GUP
effect. The Planck mass is given by $m_p=\sqrt{\hbar c/G}$. The
above implies a lower bound on the length scale \be \label{2eq2}
\Delta x \ge 2 \alpha l_p, \ee
 which means that the Planck length plays
the role of a fundamental scale. Also,
 the GUP may be used to derive the modified black hole temperature by introducing  $\Delta p$ as
 the energy (temperature) of radiated
photons. The momentum uncertainty for radiated photons can be
found to be

\be \label{3eq2} \frac{ \Delta x}{2 \alpha^2 l_p^2}
  \Bigg[1- \sqrt{1-\frac{4\alpha^2 l_p^2}{(\Delta
 x)^2}}\Bigg] \le  \frac{\Delta p}{\hbar} \le \frac{ \Delta x}{2\alpha^2 l_p^2}
  \Bigg[1+ \sqrt{1-\frac{4 \alpha^2 l_p^2}{(\Delta
 x)^2}}\Bigg].
 \ee
 The left inequality provides small corrections to the
 Heisenberg's uncertainty principle for $\Delta x\gg \alpha l_p$ as $\Delta p \ge
 \hbar/\Delta x +\hbar \alpha^2l_p^2/(\Delta x)^3+\cdots$~\cite{Park}.
On the other hand, the right inequality implies that $\Delta p$
cannot be arbitrarily large in order that the GUP in (\ref{1eq2})
makes sense.

For simplicity we use the Planck units of $c=\hbar=G=k_B=1$ which
imply that $l_p=m_p=1$ and $\beta=\alpha^2$. Considering the GUP
effect on the near-horizon and $\Delta x=r_{+}=2m$ from
Eq.(\ref{smet})~\footnote{By modelling a black hole as a black box
with linear size $r_+$, the uncertainty in the position of an
emitted particle by the Hawking radiation may take $\Delta x \sim
r_+$ with $r_+$ the radius of event horizon. In the absence of the
GUP effect, the horizon radius is given by $r_+=2m$. In the
presence of the GUP effect, we do not know a precise form of the
horizon radius $r_+=r_+(m,\beta)$ unless a GUP-corrected metric is
known. However, we assume that $\Delta x \sim r_+$ would be valid
even with the GUP  effect. }, the
 relation (\ref{3eq2})  reduces to

\be \label{4eq2} m
  \Bigg[1- \sqrt{1-\frac{\beta }{m^2}}\Bigg] \le  \beta \Delta p \le
  m
  \Bigg[1+ \sqrt{1-\frac{\beta}{m^2}}\Bigg].
 \ee
Replacing $\beta$ with $1/2\omega$, the above leads to a relation
\be \label{analogy} r_-\le \frac{\Delta p}{2  \omega} \le r_+.
 \ee
Here we wish to mention that a replacement of  $\beta\to
1/2\omega$ was done because both sides of Eq.(\ref{4eq2}) have
mathematically the same form as Eq.(\ref{sol}). It seems that
Eq.(\ref{analogy}) indicates a connection between quantum and
classical properties of black holes in the deformed
Ho\v{r}ava-Lifshitz gravity.

At first sight, it is unclear why the GUP effect on the
Schwarzschild black hole is related to   black holes in the
deformed Ho\v{r}ava-Lifshitz gravity. However,  it was known that
the generalized uncertainty principle provides naturally a UV
cutoff to the local quantum field theory as gravity
effects~\cite{CMOK,KLM}. The GUP relation of Eq.(\ref{crgup}) has
an effect on the density of states in momentum space as
\begin{equation} \label{dgup}
\frac{d^3\vec{p}}{(1+\beta \vec{p}^2)^3}
\end{equation}
with an important factor of $1/(1+\beta \vec{p}^2)^3$, which
effectively cuts off the integral beyond $p=1/\sqrt{\beta}$. We
wish to mention that this factor may be related to the Cotton-term
of $C_{ij}C^{ij}$ in Eq.(\ref{action2})  because the latter
contains  a six-order derivative.

The right-hand side of Eq.(\ref{crgup}) includes a
$\vec{p}$-dependent term  and thus affect the cell size in phase
space as ``being $\vec{p}$-dependent". Making use of the Liouville
theorem, one could show that the invariant weighted phase space
volume under time evolution is given by
\begin{equation}
\frac{d^3\vec{x}d^3\vec{p}}{(1+\beta \vec{p}^2)^3},
\end{equation}
where the classical commutation relations corresponding to the
quantum commutation relation of Eq.(\ref{crgup}), namely,  $\{x_i,
p_j\}=(1+\beta p^2)\delta_{ij}$, $\{p_i,p_j\}=0$, and
$\{x_i,x_j\}=2\beta(p_ix_j-p_jx_i)$, are used. Actually, $\beta
(=\alpha^2)$ plays the role of a  UV cutoff of the consequent
momentum integration~\cite{KLM}.

In order to make  a further connection between the GUP and the
Ho\v{r}ava-Lifshitz gravity, we calculate the GUP-corrected
temperature.  If the mass of the emitted particle is neglected,
the uncertainty in the energy of the emitted particle is given by
\be \Delta E \sim \Delta p. \ee By considering that $\Delta E$,
which could be identified with the characteristic temperature of
the Hawking radiation, saturates the left inequality in
Eq.(\ref{analogy}), one has the Hawking temperature \be
T_{GUP}=\frac{ r_+}{8\pi
\beta}\Bigg[1-\sqrt{1-\frac{4\beta}{r_+^2}}\Bigg]. \ee Here a
calibration factor ``$1/4\pi$" was introduced to have agreements
with the usual Hawking temperature of the Schwarzschild black hole
in the leading term, for a large black hole with $r_+\gg \beta$:
\be \label{gupasy}  T_{GUP}\simeq \frac{1}{4\pi}
\Bigg[\frac{1}{r_+}+\frac{\beta}{r_+^3}+\cdots\Bigg]. \ee On the
other hand, from Eq.(\ref{temp}) we have temperature for large
Ho\v{r}ava-Lifshitz black holes \be\label{hlasy} T \simeq
\frac{1}{4\pi} \Bigg[\frac{1}{r_+}-\frac{3}{2\omega
r_+^3}+\cdots\Bigg]. \ee Comparing Eq.(\ref{gupasy}) with
Eq.(\ref{hlasy}) leads to a difference that the GUP effect
($\beta$) with the minimum length $2 \sqrt{\beta} l_p$  always
increases the Hawking temperature, while the Ho\v{r}ava-Lifshitz
gravity $\omega$ decreases the Hawking temperature and it arrives
at $T=0$. Furthermore, we could not confrim the replacement of
$\beta \to 1/2\omega$ on the temperature side. Thus, the relation
of Eq.(\ref{analogy}) is still obscure.

We have studied thermodynamics of black holes in the deformed
Ho\v{r}ava-Lifshitz gravity.
  For $\lambda =1$, the black hole behaves like the
  Reissner-Nordstr\"om black hole  because of the extremal point with $m=m_e,~T=0,~C=0$ and
  the maximum point  as the Davies' point with
  $T=T_m,~C_m=[\infty\to -\infty]$~\cite{Myung}.  Hence, for small black hole,
  it is  quite different from the Schwarzschild black
  hole of Einstein gravity even though  the Schwarzschild black
  hole is recovered  for large
  $r_+$. Explicitly, there is no such points of $r_+=r_e,r_m$ in the
  Schwarzschild black hole.
In this work, we did not obtain  the entropy of the deformed
Ho\v{r}ava-Lifshitz black holes. However, we  would like to
mention that either the Wald formalism  or the entropy function
formalism may  be applied to the Ho\v{r}ava-Lifshitz action to
find the entropy correctly because this action contains higher
order curvature terms like $R^2$ and $R_{ij}R^{ij}$.

Finally, we could not confirm a solid connection between the GUP
and Ho\v{r}ava-Lifshitz gravity, although we have obtained partial
connections between them.

\section*{Acknowledgement}
The author  was supported by the SRC Program of the KOSEF through
the Center for Quantum Spacetime (CQUeST) of Sogang University
with grant number R11-2005-021-03001-0.

\end{document}